\documentclass[11pt,a4,twoside]{article}
\usepackage[dvips,a4paper,left=2.75cm,right=2.75cm,top=2.0cm,bottom=2.0cm,headsep=1em]{geometry}
\usepackage{fancyhdr}
\usepackage{titlesec}
\usepackage[latin1]{inputenc}
\usepackage{amsmath,amsfonts}
\usepackage{rotating}
\usepackage[font={small}]{caption}
\usepackage{natbib}
\usepackage{lmodern,slantsc}
\usepackage[breaklinks,
colorlinks=true, linkcolor=blue, citecolor=black, urlcolor=blue,
pdfborder={0 0 0}, pdfpagelabels]{hyperref}
\usepackage{tikz}
\usepackage{enumitem}

\def\v0{\boldsymbol{0}}

\def\L{\langle}
\def\R{\rangle}
\newlength{\FigureHeight}
\newlength{\FigureHeightHalf}

\numberwithin{equation}{section}
\titleformat{\section}
{\large\bfseries}{\thetitle.}{0.5em}{}
\titleformat{\subsection}
{\normalfont\itshape\filcenter}{\normalfont\thetitle.}{0.5em}{}
\begin{document}

\title{A note on the notion ``statistical symmetry"}
\author{M. Frewer$\,^1$\thanks{Email address for correspondence:
frewer.science@gmail.com}$\:\,$, G. Khujadze$\,^2$ \& H. Foysi$\,^2$\\ \\
\small $^1$ Tr\"ubnerstr. 42, 69121 Heidelberg, Germany\\
\small $^2$ Chair of Fluid Mechanics, Universit\"at Siegen, 57068
Siegen, Germany}
\date{{\small\today}}
\clearpage \maketitle \thispagestyle{empty}

\vspace{-2em}
\begin{abstract} \noindent
A critical review is presented on the most recent attempt to
generally explain the notion of\linebreak ``statistical symmetry".
This particular explanation, however, is incomplete and misses one
important and essential aspect. The aim of this short note is to
provide this missing information and to clarify this notion on the
basis of a few instructive examples.

\vspace{0.5em}\noindent{\footnotesize{\bf Keywords:} {\it
Statistical Physics, Dynamical Systems, Stochastic Processes,
Random Walk, PDFs, Turbulence, Lie Groups, Symmetries, Invariant
Solutions, Scaling Laws, Principle
of Causality}}$\,$
\end{abstract}

\section{Introduction\label{S1}}

An attempt to explain the notion of ``statistical symmetry" was
recently made in \cite{Oberlack15.6}. Although stated correctly
that such a name has ``been used in other areas\linebreak of
science, e.g., in the study of dynamical systems
[6]=\cite{Aubry95}, statistical physics [7]=\cite{Birman85}, or
even sensory coding [8]=\cite{Turiel03}"~[p.~1], and that ``even
if one particular field (or image) does not verify the symmetry,
it can be\linebreak observed over a large ensemble of fields (or
images)"~[p.~1], these statements still miss one\linebreak
important fact, namely to mention that if the statistical
description is based on an underlying\linebreak deterministic
theory, then every statistical symmetry must have some kind of
cause from which it can emerge, whereby the cause itself certainly
need not to be a symmetry transformation. The obvious reason is
that since the deterministic equations due to their spatially
nonlocal and temporally chaotic behavior induce the statistical
equations, and not vice versa, and since any symmetry in physics
is always defined as a transformational process, namely as an
invariant transition from one state to another, this occurrent
transition, i.e.~the transition itself must then have some
deterministic cause if a symmetry is to be observed
statistically.$\phantom{xx}$

Hence, the cause for a statistical symmetry must be encoded as a
change on the lower deterministic (fluctuating) level such that
for a large ensemble of fields (or images) a symmetry can be
observed. Important to note here again is that the cause itself
need not to be symmetry, but at least there must be an occurring
change on the fluctuating level, namely from one state to another
such that on the averaged level this change can then be observed
as a symmetry. In other words, a statistical symmetry is based on
an active construction process resulting from a (mostly
non-invariant) change of its underlying deterministic system. All
statistical symmetries derived and discussed in the above cited
studies \citep{Aubry95,Birman85,Turiel03} are of this type, in
that they all have a cause which all originate from a
non-invariant change in the underlying deterministic system --- in
particular, because they are all based on pure space-time
coordinate transformations, which not only transform the
deterministic but also the induced statistical state, however,
such that on the mean level a symmetry can be observed.

To be explicit, the study by \cite{Aubry95} investigates the
connection, i.e. the cause and effect between the (microscopic)
spatiotemporal symmetries of a space-time function and the
(macroscopic) symmetries of a statistical quantity, as~e.g.~for
the two-point correlation, while \cite{Birman85} defines
``statistical symmetries" as ``symmetries in the usual sense"
[p.~388], namely as deterministic coordinate transformations
(translations, rotations, reflections, etc.) which only get the
label `symmetry' as soon as they leave the configurational or
temporal average invariant. The third cited study by
\cite{Turiel03} considers e.g. a deterministic spatial scaling
transformation (via a wavelet transformation) in every single
image which then induces a local invariance of a fixed scale in
the ensemble of all images, i.e. the cause of this statistical
symmetry is the (non-invariant) transformation of each single
image which then induces the effect of a symmetry in the ensemble
of all images.

The same is also true for the statistical symmetry discussed in
\cite{Oberlack15.6}, which actually has been taken from
\cite{Aubry95}~[p.~794], however, just reformulated for an
unsteady Poiseuille flow in a channel. Indeed, the discussed
statistical reflection symmetry for the mean velocity profile
about the center plane $y=0$ in \cite{Oberlack15.6}~[p.~1] has a
cause, namely the spatial reflection transformation itself, which
changes the instantaneous state $U(x,y,z)$ into a different state
$U(x,-y,z)\neq U(x,y,z)$ with equal probability, such that within
an ensemble of the instantaneous velocity field this change or
transformation emerges then as a symmetry on the mean level: $\L
U\R(-y)=\L U\R(y)$.

Another simple example is to consider a fully developed turbulent
channel flow which is statistically stationary in time. Although
the time shift transformation is not a symmetry of the
instantaneous velocity field, it nevertheless serves as the cause
to emerge as a symmetry in the correspondingly induced statistical
field. To be explicit, let us view this example from the
perspective of a numerical simulation: Imagine we have a DNS
result of a fully developed turbulent channel flow. Within this
flow regime we then collect an ensemble of the instantaneous
velocity field, say, of the streamwise component at different
times, however, with a time-interval always sufficiently large in
order to ensure statistical independence. If one now takes the
ensemble average over these different instantaneous field
realizations, we observe, as an effect, that the flow statistics
is independent of time. But the cause of this effect is the
preceding process of evaluating or choosing the instantaneous
velocity field at different times, which itself just represents a
deterministic time shift transformation, in that the considered
DNS field successively gets transformed to different times in
order to obtain the elements of this particular ensemble. In other
words, the statistical time translation symmetry emerges as an
effect from the cause of transforming the underlying deterministic
(instantaneous) field to different times.

A further instructive example is discussed in the next section,
where we even present a more detailed and at the end also a more
general investigation of this cause-and-effect issue for
statistical symmetries.

Currently, within incompressible fluid mechanics, no other or new
statistical symmetries for the Euler and Navier-Stokes equations
are known yet than the classical ones listed through Eqs.~(8)-(13)
in \cite{Oberlack14.1}, which all have their origin in the
deterministic equations. In clear contrast, of course, to the {\it
newly} proposed statistical symmetries given
in~\cite{Oberlack14.1} by Eq.~(14) and Eq.~(16) for the
multi-point correlation functions, or by Eq.~(42) and Eq.~(63) for
the PDFs, which are emerging without any cause at all. As shown in
\cite{Frewer14.3,Frewer14.1}, one even runs into a contradiction
as soon as one tries to establish a cause of any kind. The problem
is that {\it all} multi-point correlation functions and {\it all}
multi-point PDFs get scaled or translated by a spatially constant
quantity for which at the same time the coordinates and all
deterministic fields stay invariant.\footnote[2]{The magnitude of
this problem also becomes apparent through the following
contradictive statement made in \cite{Oberlack15.6}: Saying on
p.~2 after Eq.~(5) that {\it ``Here, the single [deterministic]
velocity fields v and U are not rescaled [or changed] but the
share of laminar solutions in the ensemble changes. This could be
caused, e.g., by the change of initial conditions or external
disturbances"} is obviously fictitious and\linebreak unphysical.
Because, how is it possible that if initial conditions change or
external disturbances appear that then the underlying
deterministic fields do not change? This conflict in assumption
ultimately also explains their inconsistent result summarized on
p.~4: {\it ``After such [symmetry] transformations the statistics
change, although the [under\-lying] instantaneous velocities $U$
in separate realizations of the flow are not transformed (i.e.,
they are not rescaled or translated)"}, constitutes a clear
inconsistency, because again, from where should this change on the
statistical level come from when the underlying deterministic
system itself is not changing?} Hence, since no cause for those
new statistical symmetries can be constructed, they obviously
violate the classical principle of causality as we fully
elaborated in our Comment~\citep{Frewer15.C,Frewer14.3}. This
unphysical behavior can also be clearly seen when comparing to DNS
data \citep{Simens09,Borrell13}; a detailed summary of this
analysis is given e.g. in \cite{Frewer15.X}. The only conclusion
to be drawn from this negative result, is to discard these newly
proposed statistical symmetries in order to avoid any misleading
conclusions in the theory of turbulence.

\section{The causality in a statistical symmetry at the example of a
random walk\label{S2}}

Without loss of generality, we consider the spatial 1D case for
simplicity. As we know, the problem `random-walk' can be discussed
mathematically in two {\it equivalent} ways, either directly as a
dynamical process through a stochastic variable, or indirectly as
an evolutional process through a probability distribution
characterizing the dynamics of this stochastic
variable.\footnote[2]{The theory on which this section is based
can be found in any standard text book on stochastic processes,
see e.g. \cite{Gardiner85,Salinas01,Mahnke09}.} In both approaches
one has to distinguish between the discrete (fine-grained) random
walk on a space-time lattice and the continuous (coarse-grained)
random walk in a space-time continuum in which the emergent
concept of diffusion arises.

\subsection{The discrete (fine-grained) random walk}

The random walk of a spatially 1D moving particle on a space-time
lattice can be investigated

\vspace{-0.5em}
\begin{itemize}
\item[(i)] either as the recursive stochastic process
\begin{equation}
m_{N+1}=m_N+\xi_N, \label{150621:2329}
\end{equation}
where $m_N$ is the stochastic variable for finding the particle at
position $m$ after $N$ successive steps from the initial position,
say $m=0$, being forced by a random variable $\xi_N$, which in
each step $N$ can only take one of two possible values $\xi_N=\pm
1$, where for simplicity we assume that both outcomes occur with
equal probability $1/2$,
\vspace{-0.5em}
\item[(ii)] or, equivalently, by the recursive probability
(master) equation
\begin{equation}
P_{N+1}(m)=\frac{1}{2}\cdot P_N(m-1)+\frac{1}{2}\cdot P_N(m+1),
\label{15061247}
\end{equation}
to find the distribution $P_N(m)$ that after $N$ steps the
particle can be found at position~$m$.
\end{itemize}

\vspace{-0.5em}\noindent In both cases (i) and (ii) the variables
$N\geq 0$ and $m$ are integers, such that $-N\leq m\leq
N$\linebreak is always satisfied. The solution of
\eqref{150621:2329} is given by
\begin{equation}
m_N=\sum_{i=0}^{N-1}\xi_i,\label{150622:1016}
\end{equation}
with a vanishing expectation value or first moment $\L m_N\R=0$
and a variance or second moment which equals to the number of
steps performed $\L m^2_N\R=N$, due to the zero mean and due to
the statistical independence of the random variable $\xi_i$
respectively: $\L\xi_i\R=0$ and $\L \xi_i\cdot
\xi_j\R=\delta_{ij}$. The solution to equation \eqref{15061247} is
given by the binomial distribution
\begin{equation}
P_N(m)=\frac{N!}{\left(\frac{N+m}{2}\right)!\left(\frac{N-m}{2}\right)!}
\cdot\left(\frac{1}{2}\right)^N, \label{15061414}
\end{equation}
which only gives a physical solution to the random walk if
$N_\pm=\frac{N\pm m}{2}\in\mathbb{N}_0$, where $N_+$ is the number
of steps to the right and $N_-=N-N_+$ the number of steps to the
left in order to end after $N$ steps at position $m$. The solution
\eqref{15061414} is already properly normalized
\begin{equation}
\sum_{N_+=0}^N \frac{N!}{N_+!
(N-N_+)!}\left(\frac{1}{2}\right)^N=1,
\end{equation}
and gives the same values for the moments as solution
\eqref{150622:1016}: $\L m\R=2\L N_+\R-N=0$ and $\L m^2\R=\L
(2N_+-N)^2\R=N$. Finally, let us introduce, instead of the
non-dimensional integer variables $N$ and $m$, the real physical
variables $t$ and $x$ respectively:
\begin{equation}
t=N\cdot \tau, \qquad x=m\cdot l,
\end{equation}
where $\tau$ is a constant time interval between {\it all}
successive steps of the same length $l$. That means, $P_N(m)$
\eqref{15061414} can then be interpreted as the probability of
finding the particle at the physical position $x=m\cdot l$ at the
time $t=N\cdot\tau$. In the following, the parameters $l$ and
$\tau$ are arbitrary but fixed (finite) values. Now, let's
consider the following point transformation of variables
\begin{equation}
\tilde{t}=c^2\cdot t,\quad \tilde{x}=c\cdot x,\quad \tilde{P}=P,
\quad\text{where}\;\; c\in\mathbb{Z}/\{0\}, \label{1506:1540}
\end{equation}
which induces the transformation in the original variables as
\begin{equation}
\tilde{N}=c^2\cdot N,\quad \tilde{m}=c\cdot m,\quad \tilde{P}=P.
\end{equation}
It is obvious that it's {\it not} a symmetry transformation of
equations \eqref{150621:2329} and \eqref{15061247}, since neither
solution \eqref{150622:1016} nor solution \eqref{15061414} gets
mapped to a new solution of equation \eqref{150621:2329} and
equation \eqref{15061247} respectively, i.e. the mapped function
of solution \eqref{150622:1016}
\begin{equation}
\tilde{m}_{\tilde{N}}\;\;\mapsto\;\; m_N=\frac{1}{c}\cdot
\sum_{i=0}^{c^2\cdot N-1}\xi_i,
\end{equation}
as well as the mapped function of solution \eqref{15061414}
\begin{equation}
\tilde{P}_{\tilde{N}}(\tilde{m})\;\;\mapsto\;\;
P_N(m)=\frac{(c^2\cdot N)!}{\left(\frac{c^2\cdot N+c\cdot
m}{2}\right)!\left(\frac{c^2\cdot N-c\cdot m}{2}\right)!}
\cdot\left(\frac{1}{2}\right)^{c^2\cdot N},
\end{equation}
are no longer solutions of equations \eqref{150621:2329} and
\eqref{15061247} anymore.

\subsection{The continuous (coarse-grained) random walk}

The limit of large $N$ and $m$, i.e. $N\gg \tau$ and $m\gg l$, is
to be seen as a coarse graining process of the random walk. In
general, coarse-graining is understood as a process which
eliminates the ``uninteresting" fast and small variables and keeps
the coarse-grained variables with time and space scales much
larger than the microscopic (fined-grained) scales $\tau$ and $l$.
And exactly this is what this limit of large $N$ and $m$ does, for
which then (i) the discrete (fine-grained) stochastic
process~\eqref{150621:2329} turns into the continuous
(coarse-grained) 1D Wiener process
\begin{equation}
dx_t=\sqrt{2D}\cdot dw_t, \label{150622:1146}
\end{equation}
with its well-known statistical properties $\L w_t\R=0$ and $\L
w_t^2\R=t$, and for which (ii) the discrete (fine-grained)
probability distribution equation \eqref{15061247} turns into the
associated continuous (coarse-grained) Fokker-Planck equation
\begin{equation}
\frac{\partial P}{\partial t}=D\cdot\frac{\partial ^2P}{\partial
x^2}. \label{1506:1543}
\end{equation}
Both equations \eqref{150622:1146} and \eqref{1506:1543} describe
the physical process of diffusion in a mathematical equivalent
manner, where $D=l^2/2\tau$ denotes the diffusion coefficient.

\noindent Suddenly, in the limit of large $N\gg \tau$ and $m\gg
l$, i.e. in the process of coarse-graining the defining system,
the transformation \eqref{1506:1540}, which keeps its validity as
a variable transformation in this limit, turns into a {\it
symmetry} transformation for the continuous (coarse-grained)
equations \eqref{150622:1146} and \eqref{1506:1543}. That means,
{\it any} solution of~\eqref{1506:1543} gets mapped into a new
solution. For example, the properly normalized solution
\begin{equation}
P(t,x)=\frac{1}{\sqrt{4\pi D t}}\cdot e^{-\frac{x^2}{4Dt}},
\label{150622:2210}
\end{equation}
which satisfies the initial condition $P(0,x)=\delta(x)$ as a
Cauchy problem, gets mapped by transformation \eqref{1506:1540} to
a new solution of \eqref{1506:1543}
\begin{equation}
\tilde{P}(\tilde{t},\tilde{x})\;\;\mapsto\;\;
P(t,x)=\frac{1}{\sqrt{4\pi D \cdot c^2\cdot t}}\cdot
e^{-\frac{x^2}{4Dt}},
\end{equation}
satisfying a different, the transformed initial condition
$\tilde{P}(0,\tilde{x})\mapsto P(0,x)=\delta(x)/|c|$ with the
normalizing constant~$|c|$.~And if, in addition, the symmetry
coordinate transformation~\eqref{1506:1540} would be augmented by
the appropriate symmetry scaling in the dependent variable $P$ as
\begin{equation}
\tilde{t}=c^2\cdot t,\quad \tilde{x}=c\cdot x,\quad
\tilde{P}=\frac{1}{|c|}\cdot P, \label{150622:2212}
\end{equation}
then the Cauchy initial-value solution \eqref{150622:2210} even
turns into a {\it self-similar} solution under the combined
scaling symmetry \eqref{150622:2212} admitted by equation
\eqref{1506:1543}. Also the general solution
\begin{equation}
x_t=x_0+\sqrt{2D}\cdot w_t, \label{150622:2257}
\end{equation}
of its associated stochastic differential equation
\eqref{150622:1146} forms a {\it self-similar} solution under the
scaling symmetry \eqref{1506:1540}, if the initial condition is
placed at the origin: $x_0=0$. To explicitly see that the
coordinate transformation \eqref{1506:1540} is admitted as a
symmetry by equation \eqref{150622:1146}, or that
\eqref{150622:2257} is a self-similar solution under this scaling
if $x_0=0$, it is necessary to realize that the defining time
transformation $\tilde{t}=c^2\cdot t$ induces the following
scaling in the Wiener process (see e.g.
\cite{Gaeta99,Srihirun07,Abdullin14})
\begin{equation}
w_{\tilde{t}} = c\cdot w_t=:\tilde{w}_{\tilde{t}},
\end{equation}
that means, unlike the discrete (fine-grained) random walk, the
continuous (coarse-grained) random walk is scale invariant, i.e.
if $w_t$ is a Wiener process then $c^{-1}w_{c^2t}$ is again a
Wiener process.

\subsection[Conclusion]{Conclusion\footnote[2]{To avoid an unnecessary overload of
formula referencing in the text and to therefore obtain a better
readability, we will base the conclusion only on the probability
distribution equations \eqref{15061247} and \eqref{1506:1543} of
the random walk. It is obvious that this conclusion will then also
hold identically for the associated and mathematically equivalent
stochastic processes \eqref{150621:2329} and \eqref{150622:1146},
respectively.}}

Although transformation \eqref{1506:1540} is a symmetry
transformation only on the coarse-grained level \eqref{1506:1543}
and not on the fine-grained level \eqref{15061247}, the
transformation \eqref{1506:1540} itself is nevertheless a valid
variable transformation on both levels. And since the combined
system \eqref{15061247} \& \eqref{1506:1540}\linebreak on the
fine-grained level uniquely induces the corresponding system
\eqref{1506:1543} \& \eqref{1506:1540} on the coarse grained
level, and not vice versa, we have a working relation of cause and
effect:\linebreak During the coarse-graining process (i.e. in the
limit of large $N\gg \tau$ and $m\gg l$) the variable
transformation \eqref{1506:1540} on the fine-grained level (the
cause) maintains to be a variable transformation also on the
coarse-grained level (the effect), but which, as a pleasant side
effect of this coarse-graining process, turns out to be
additionally a {\it symmetry} transformation (which in this simple
case, of course, has the exact same structure as the underlying
transformation \eqref{1506:1540}). In other words, the statistical
symmetry on the coarse-grained level (the
$\hspace{0.025cm}$effect)\linebreak is linked to an underlying
variable transformation on the fine-grained level (the cause),
which itself, obviously, need {\it not} to be a symmetry.

\section{The conclusion in the general case}

It's obvious that the above specific conclusion can be fully
generalized: Formally, let $F$ be any deterministic system, and
let $\L\cdot\R$ denote any coarse-graining or averaging process.
Then we can make the following implications:
\begin{align*}
\text{{\bf I.1:}} & & \text{{\it Cause:}\, Let $\mathsf{T}$ be any
variable transformation of $F$, i.e.
$\mathsf{T}F=\tilde{F}$}\hspace{2.4cm}\\[0.25em]
&\!\!\!\!\!\!\! \Rightarrow & \text{{\it Effect:}\,
$\mathsf{T}^*:=\L\mathsf{T}\R$ is a variable transformation of
$F^*:=\L F \R$, i.e. $\mathsf{T}^*F^*=\tilde{F}^{*\prime}$},\,
\end{align*}

\noindent and if further the variable transformation commutes with
the coarse-graining operator, i.e. $\mathsf{T}^*\L F\R=\L
\mathsf{T}F\R$, then we have the additional entangled relation
$\tilde{F}^*=\L \tilde{F}\R =\tilde{F}^{*\prime}$. Amongst many
possible transformations, this is in particular the case, if
$\mathsf{T}=\mathsf{S}$ is a symmetry transformation, for which we
then have the implication\footnote[2]{Note that we only consider
here the admitted symmetries of equations, and not of their
solutions. Although a symmetry of an equation maps the set of {\it
all} its solutions into itself, i.e., although one solution gets
mapped into another solution, this symmetry, however, is not
automatically admitted by a {\it particular} solution of this
equation when emerging from specific initial or boundary
conditions. The presence of such external conditions may not be
compatible with the equations' symmetry, thus giving rise to
solutions which do not reflect the symmetry of the equation
itself. In the worst case, this symmetry is not even reflected in
some asymptotic regime of the solution, which then is also known
as the effect of spontaneous symmetry breaking.}
\begin{align*}
\text{{\bf I.2:}} & & \text{{\it Cause:}\, Let $\mathsf{S}$ be a
symmetry transformation of $F$, i.e.
$\mathsf{S}F=F$}\hspace{2.6cm}\\[0.25em]
&\!\!\!\!\!\!\! \Rightarrow & \text{{\it Effect:}\,
$\mathsf{S}^*:=\L\mathsf{S}\R$ is a symmetry transformation of
$F^*:=\L F \R$, i.e. $\mathsf{S}^*F^*=F^{*}$},
\end{align*}

\noindent where, for example, the relation $\mathsf{S}^*F^*=F^*$
can be used as a natural closure restriction, if the
coarse-grained system $F^*$ is unclosed. Now, it is obvious that
the inverse or opposite conclusion of I.2 is false: A symmetry
$\mathsf{S}^*$ on the coarse-grained level does not necessarily
emerge from a symmetry $\mathsf{S}$ on the fine-grained level, as
it was already shown in the above example of the diffusion
equation and its underlying discrete random walk. But, since the
inverse conclusion of I.1 is true, we have the combined
implication
\begin{align*}
\text{{\bf I.3:}} & & \text{{\it Effect:}\, Let
$\mathsf{\Sigma}^*$ be a symmetry transf. of
$F^*:=\L F \R$, i.e. $\mathsf{\Sigma}^*F^*=F^{*}$}\hspace{1.7cm}\\[0.25em]
&\!\!\!\!\!\!\! \Rightarrow & \text{{\it Cause:}\, At least one
variable transf. $\mathsf{Q}$ of $F$ must exist, such that $\L
\mathsf{Q}\R=\mathsf{\Sigma}^*$}.\;\:
\end{align*}

\noindent Hence, if such a variable transformation of {\it any}
kind $\mathsf{Q}$ {\it cannot} be constructed, we have a violation
of cause and effect, because we are dealing with a specific
variable transformation $\mathsf{\Sigma}^*$ on the coarse-grained
level which cannot be constructed from {\it any} kind of variable
transformation on the fine-grained level. That means there is an
effect without a cause, which is not physical. And exactly this is
the case for the new statistical symmetries in
\cite{Oberlack14.1}, which constitutes the objection {\it ``I.
Violation of the causality principle"} in our Phys.-Rev.-E-Comment
\citep{Frewer15.C}.

In general it is difficult to find or to construct a certain
variable transformation $\mathsf{Q}$ on the fine-grained level
such that its coarse graining turns the transformation into the
given symmetry $\mathsf{\Sigma}^*=\L\mathsf{Q}\R$. This is an
inverse problem, for which in general no analytical solution in
closed form may exist. But since all new statistical symmetries of
\cite{Oberlack14.1}, which were first determined and introduced in
\cite{Oberlack10}, are of a globally trivial form, it is
analytically easy to show that no variable transformation of any
kind on the fine-grained level in the sense of $\mathsf{Q}$ can be
constructed at all. Every approach to construct such a
transformation leads to a clear contradiction, as it was
independently shown not only for the Lundgren-Monin-Novikov
equations (see \cite{Frewer15.C,Frewer14.3}), but also for the
Hopf equation (see Section 4.1 in \cite{Frewer14.1}), as well as
for the multi-point correlation equations (see Sections 4.2 \& 4.3
in \cite{Frewer14.1}). In particular for the new statistical
scaling symmetry of the multi-point equations (Eq.$\,$(16) in
\cite{Oberlack14.1}), it is straightforward to show that it's
already by definition impossible to construct any kind of variable
transformation for the instantaneous (fluctuating) velocity field
such that it can induce this symmetry on the mean (averaged)
level, where {\it all} velocity correlation functions
$\mathsf{H}_{\{n\}}$ are just multiplied by the {\it same} global
factor $e^{a_s}$ (see the proof e.g. in Appendix~A of
\cite{Frewer14.3}). The same issue we face for the coarse-grained
multi-point PDFs (Eq.$\,$(63) in \cite{Oberlack14.1}), where again
{\it all} PDFs get scaled by this very same constant factor
$e^{a_s}$, which, as before, simply cannot be linked to an
underlying fine-grained variable transformation of {\it any} kind,
and, therefore, inherently violates the classical principle of
causality since there would be an effect without a cause. The same
is also true for the new statistical translation symmetry, where
{\it all} multi-point PDFs (Eq.$\,$(42) in \cite{Oberlack14.1})
get translated by the {\it same} spatially global shift~$\psi$.

Both these two new statistical symmetries are simply unphysical,
as are also their consequences when trying to generate statistical
scaling laws from them. As already said in the introduction, their
unphysical behavior becomes clearly visible when comparing to DNS
data \citep{Simens09,Borrell13}: Those scaling laws which involve
any parameters from the unphysical symmetries, are completely
unable to predict the correct DNS multi-point correlations
behavior beyond the second order \citep{Frewer14.1}. But, as soon
as these unphysical group parameters are removed, the matching to
the DNS data improves by several orders of magnitude and becomes
well-defined again for all orders~$n$, which definitely is a
further and (from the analytical investigation) independent
indication that the new statistical symmetries presented in
\cite{Oberlack14.1} are unphysical
--- here we refer to \cite{Frewer15.X} where we summarized our
conclusion in more detail. Hence, as said before, these symmetries
need to be discarded in our opinion, as they will only
unnecessarily lead to erroneous conclusions in the theory of
turbulence.

\bibliographystyle{jfm}
\bibliography{BibDaten}

\end{document}